# Structural transitions in dense disordered silicon from quantum-accurate ultra-large-scale simulations

Volker L. Deringer,[1,]* Noam Bernstein,[2] Gábor Csányi,[3] Mark Wilson,[4]
David A. Drabold[5] & Stephen R. Elliott[6]

[1]*Department of Chemistry, Inorganic Chemistry Laboratory, University of Oxford, Oxford OX1 3QR, United Kingdom*

[2]*Center for Materials Physics and Technology, U.S. Naval Research Laboratory, Washington, DC 20375, United States*

[3]*Engineering Laboratory, University of Cambridge, Cambridge CB2 1PZ, United Kingdom*

[4]*Department of Chemistry, Physical and Theoretical Chemistry Laboratory, University of Oxford, Oxford OX1 3QZ, United Kingdom*

[5]*Department of Physics and Astronomy, Ohio University, Athens, Ohio 45701, United States*

[6]*Department of Chemistry, University of Cambridge, Cambridge CB2 1EW, United Kingdom*

*volker.deringer@chem.ox.ac.uk

**Structurally disordered materials continue to pose fundamental questions,[1–4] including that of how different disordered phases ("polyamorphs") can coexist and transform from one to another.[5–7] As a widely studied case, amorphous silicon ($a$-Si) forms a fourfold-coordinated, covalent random network at ambient conditions, but much higher-coordinated, metallic-like phases under pressure.[8–10] However, a detailed mechanistic understanding of the liquid–amorphous and amorphous–amorphous transitions in silicon has been lacking, due to intrinsic limitations of even the most advanced experimental and computational techniques. Here, we show how machine-learning (ML)-driven simulations can break through this long-standing barrier, affording a comprehensive, quantum-accurate, and fully atomistic description of all relevant**

**liquid and amorphous phases of silicon. Combining a model system size of 100,000 atoms (ten-nanometre length scale) with a prediction accuracy of a few meV per atom, our simulations reveal a remarkable, three-step transformation sequence for *a*-Si under increasing external pressure. First, up to 10–11 GPa, polyamorphic low- and high-density amorphous (LDA and HDA) regions are found to coexist, rather than appearing sequentially. Then, we observe a structural collapse into a distinct, very-high-density amorphous (VHDA) phase at 12–13 GPa, reminiscent of the dense liquid but being formed at a much lower temperature. Finally, our simulations indicate the transient nature of this VHDA phase: it rapidly nucleates crystallites at 13–16 GPa, ultimately leading to the formation of a poly-crystalline, simple-hexagonal structure, consistent with experiments[11–13] but not seen in earlier simulations.[9,14–16] These results shed new light on the structural nature of the archetypal disordered material system, *viz.* liquid and amorphous states of silicon, and the insight gained herein can map onto other polyamorphic network systems, such as liquid water and amorphous ices.[17,18] In an even wider context, our work exemplifies a new, ML-driven approach to multiscale materials modelling, in which predictive simulations move beyond small, simplified models and reveal phenomena on physically realistic length and time scales.**

The state-of-the-art in understanding structurally complex materials, including liquid and amorphous matter, has been reached in no small part by means of computer simulations. Still, disordered phases have been one of the most persistent challenges for simulations, requiring transferable interaction models (that are valid for all relevant structural and bonding environments), large system sizes, and long simulation times. ML-driven interatomic potentials are an emerging and powerful approach to address this challenge,[19–21] with pressure-induced transitions between *crystalline* phases of silicon having been among the very first applications of these methods,[22] and crystal nucleation in liquid silicon among the most recent ones.[23] We have previously carried out pilot studies of amorphous silicon (*a*-Si) based on molecular-dynamics (MD) simulations with a Gaussian approximation potential (GAP) ML model,[24,25] using system sizes between 512 and 4,096 atoms, and considering only the ambient-pressure regime at that time.[26,27] In the present work, we now report a landmark step forward: we performed much more extensive GAP-MD simulations of a system containing 100,000 silicon atoms (Fig. 1a), and we surveyed this system throughout the full parameter space of relevant temperatures and pressures (Fig. 1b). Comprising several million individual timesteps at this system size, such simulations would previously have only been possible with much simpler, empirically parameterised force fields of (necessarily) limited accuracy and transferability.[28,29]

High-pressure forms of matter show diverse and often unexpected atomic environments, and therefore structural exploration at gigapascal pressures requires predictive, quantum-accurate methods.[30] We therefore begin by justifying the ML potential model that we use: we developed it, based on an established GAP for silicon[25] ("GAP-18"), but now include iteratively sampled liquid and amorphous structures under very high pressures in the reference database ("GAP-18P"), as detailed in the Methods section. Uncertainty quantification, based on the Gaussian process variance,[25] shows that our ML approach can indeed reach a quantum-mechanical level of prediction accuracy, which we take to be within a few meV per atom (Fig. 1c). Beyond numerical verification, a new ML potential must also be amenable to validation against experiments:[31] here, existing X-ray diffraction data for the dense liquid[32] provide a suitable benchmark, and indeed the measured and computed structure factors, $S(q)$, are in excellent agreement over the entire pressure range up to 23 GPa (Fig. 1d). Having validated our method for the highly complex liquid phases, we may now apply it to further structural questions with full confidence.

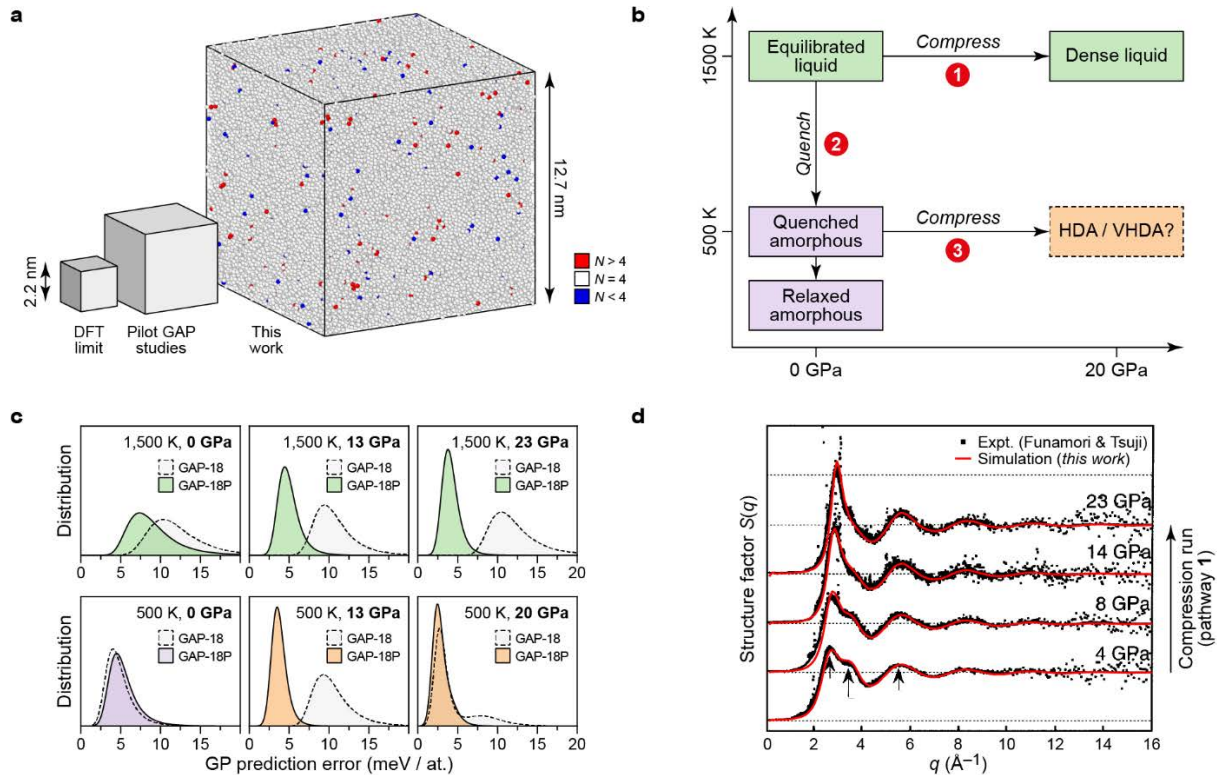

**Figure 1: Machine-learning-driven materials modelling beyond the nanometric length scale.**
(**a**) A fully relaxed amorphous silicon (*a*-Si) structure with 100,000 atoms, created in the present work. The smaller boxes on the left show the size of a 512-atom system from a recent study,[26] marking the limit of current DFT methods for simulations over several nanoseconds, and that of a 4,096-atom system in our pilot GAP-MD studies;[26] all boxes are drawn to scale. (**b**) Overview of the present study, illustrating the parameter space of temperatures and pressures in which we probe all relevant transition pathways ("1–3"). The existence and nature of (very) high-density amorphous phases have been widely discussed; they are indicated here by a dashed box and are studied in this work. (**c**) Uncertainty quantification for our ML model, using predicted errors based on the Gaussian process variance, shown for the general-purpose GAP-18 (*dashed lines*)[25] and for our new, extended GAP-18P model (*solid lines*); see Methods. (**d**) Structure factors of liquid silicon during compression (pathway 1). Computed values from our simulations (*red*) are overlaid on the experimental reference data (*black*); the figure is adapted from the original work by Funamori and Tsuji.[32] Reprinted figure with permission from Ref. 32 (https://dx.doi.org/10.1103/PhysRevLett.88.255508). Copyright 2010 by the American Physical Society. Arrows indicate the location of the maxima and a shoulder in the first peak, the latter being gradually diminished at higher pressure, all of which are correctly described by our simulations.

The first mechanism to be studied here in atomistic detail is the liquid–amorphous transition. Cooling liquid silicon at a properly chosen rate yields a high-quality, glassy *a*-Si network, as we have previously established for small GAP model structures.[26,27] We have now carried out a quench simulation for our 100,000-atom system, reducing the temperature at a rate of $10^{11}$ K s$^{-1}$ in the relevant region (Fig. 2a). The large system size and (relatively) slow cooling allow us to pinpoint the transition from a high-density liquid (HDL) to a low-density amorphous (LDA) phase, as the volume increased by about 10% between 1,195 and 1,175 K (Fig. 2a). While our initial, 1,500 K liquid appeared to be fully disordered (Fig. 2b), we observed an onset of spatial heterogeneity ("patchiness") during cooling, shown at 1,195 K, just before the transition set in. At this stage, regions of high coordination numbers (red in Fig. 2b) coexisted with others that were much closer to fourfold, "diamond-like" coordination (white), and spatial fluctuations occurred on the length scale of a few nanometres. Upon further cooling (1,195 → 1,175 K), we then observed a rapid transition to a largely fourfold coordinated, glassy network, concomitant with a sudden drop in the atomic mobility (as monitored by the mean-square displacement, MSD; Fig. 2a). Beside the coordination numbers, the overall, short- to medium-range structural similarity to crystalline silicon also increased sharply during the transition: we measure that using the Smooth Overlap of Atomic Positions (SOAP) kernel,[33] which defines a quantitative value for the structural similarity to diamond-type silicon, ranging between zero and one for each atom (Fig. 2c).[27] We may finally link the evolution of the spatial (and purely structural) heterogeneity with that of local energetic stability: "machine-learned" atomic energies, $\varepsilon_{ML}$, derived from the GAP regression model, can serve as an indicator for the stability of individual atomic environments in liquid and amorphous silicon.[27] Those regions that are low-coordinated (white in Fig. 2b) and similar to diamond-type silicon (light green in Fig. 2c) also have low – that is, favourable – ML atomic energies (blue in Fig. 2d), and vice versa.

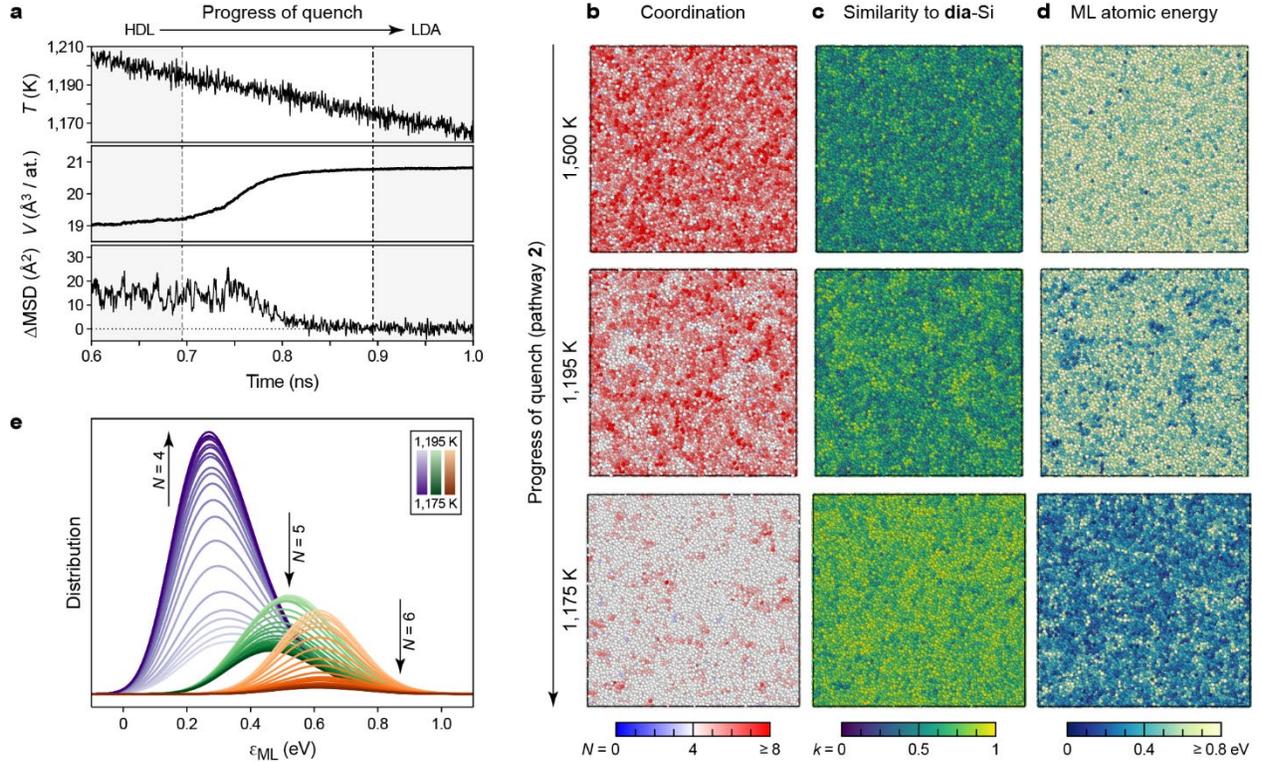

**Figure 2: Heterogeneity during quenching from the melt. (a)** Evolution of the temperature, $T$, the cell volume, $V$, and the change in atomic mean-square displacement, $\Delta$MSD (obtained by subtracting a 10-step moving average) in the relevant region of the simulation trajectory. **(b)** Structural snapshots during the quenching simulation (pathway 2 in Fig. 1b), taken from the equilibrated liquid (*top*), just before (*middle*), and just after the structural transition (*bottom*). Simulation cells are shown in plan view, offering the same perspective in all panels. Atoms are drawn as opaque spheres, and so the slice thickness is a few Å at most. Coordination numbers, $N$ (spatial cut-off = 3.1 Å), are indicated by colour coding. **(c)** Same for the SOAP-kernel similarity to ideal diamond-type crystalline ("**dia**") Si.[34] **(d)** Same for the ML atomic energies, $\varepsilon_{ML}$ (referenced to **dia**-Si).[27] **(e)** Evolution of $\varepsilon_{ML}$ shown as kernel-density estimates ("smoothed histograms"), evaluated at 1 K temperature increments between 1195 and 1175 K, and shown separately according to coordination numbers, $N$. The arrows indicate the direction of evolution of the curves with decreasing temperature, *i.e.* during the quench from the liquid to the amorphous state.

Remarkably, the distribution of $\varepsilon_{ML}$ and its evolution during the HDL → LDA transition (between 1,195 and 1,175 K) can be deconvoluted into distinct contributions from four-, five-, and sixfold bonded environments (Fig. 2e). This approach complements the colour-coded plots in Fig. 2d by now giving insight into the entirety of the system – collecting local information for a total of 21 simulation snapshots, or 2.1 million distinct atomic environments.

The second mechanism, and perhaps the most intriguing question in the context of the present work, concerns the structural transformations of *a*-Si under high pressure. Diamond-anvil cell (DAC) experiments have indicated an amorphous–amorphous transition upon compressing *a*-Si to several gigapascals, evidenced by the sudden disappearance of high-frequency Raman-peak finger-prints and by a concomitant sharp increase of the electrical conductivity (a semiconductor–metal transition), both indicative of a major change in atomistic structure.[8–10] Increasing the pressure even further, to about 14 GPa, was seen to induce crystallisation (thereby demarcating the phase boundary of dense disordered silicon),[11,12] although the experimental results may depend on the nature, origin, and purity of the sample,[13,35] and the mere appearance of Bragg peaks does not explain the mechanism by which a crystalline phase forms. Furthermore, while experiments made it possible to identify the transition in the first place, they can lend relatively little insight into the atomistic structure of the amorphous high-density phase(s). Over the years, computer simulations have led to predictions of various high-pressure structures, including those with $N = 5$ predominantly[9,10,15] and those with much higher coordination numbers,[14] presumably depending on the computational method used. No simulation has been able to reproduce the pressure-induced crystallisation of simple hexagonal (*sh*) silicon, to the best of our knowledge. Motivated by these observations (and outstanding questions), we have now carried out ML-driven simulations of our 100,000-atom *a*-Si system under isothermal compression. Hydrostatic pressure was applied at a constant rate of 0.1 GPa ps$^{-1}$, while the temperature was held at 500 K: high enough to overcome local energy barriers, but low enough to prevent melting.

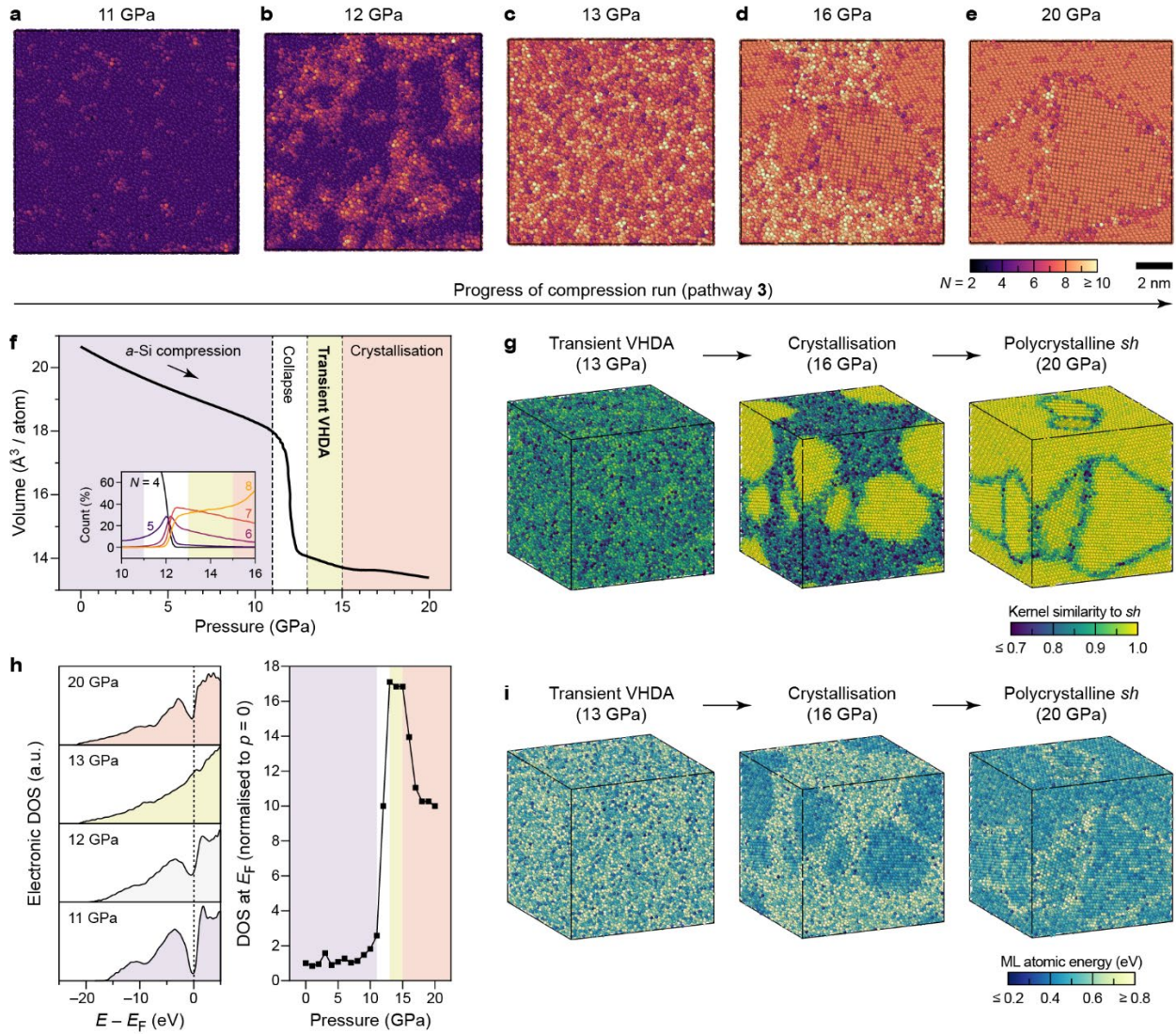

**Figure 3: Amorphous silicon at high and very high pressure.** (**a**–**e**) Structural snapshots during the isothermal compression run at 500 K (pathway 3 in the sketch in Fig. 1b), showing the coexistence of LDA-like ($N = 4$) and HDA-like ($N > 4$) regions up to 11 GPa, the collapse into a transient VHDA phase ($N \gg 4$) at 12–13 GPa, and finally the formation of simple-hexagonal (*sh*) crystallites. Colour coding indicates coordination numbers, $N$ (spatial cut-off = 2.85 Å). (**f**) Volume versus pressure during this simulation. The transition pressure, as well as the onset of crystallisation (indicated by dashed lines), are consistent with experimental reports;[11] see text. The inset shows the evolution of coordination numbers, $N$, during the structural transitions. (**g**) SOAP kernel similarity to simple hexagonal (*sh*) silicon. This analysis shows the system at 13 GPa to be fully disordered on the atomic scale and homogeneous on the nanometric scale. In contrast, *sh*-like crystallites have formed at 16 GPa, leading to nm-scale inhomogeneity. (**h**) Electronic densities of states (DOS) of the 100,000-atom model from tight-binding computations for points along the compression isotherm (*left*), and the DOS at the Fermi level, $E_F$, as a simplified measure for metallicity (*right*); see text. (**i**) ML atomic energies, $\varepsilon_{ML}$, indicating stabilisation of the *sh*-like regions.

The evolution of our *a*-Si system with increasing pressure is visualised in Fig. 3a–e, which reveals multiple interesting phenomena. Up to 11 GPa, most atoms remained in fourfold-coordinated (LDA-like) environments. However, regions of higher coordination emerged (magenta in Fig. 3a), consistent with the notion of a "high-density amorphous" (HDA) phase. A striking result is the *co-existence* of LDA- and HDA-like regions at the same temperature and pressure; that is, the simulations directly indicate the presence of polyamorphism. Being able to capture this phenomenon at all requires system sizes beyond the nanometric length scale (Fig. 1a), and the quantum-level accuracy of the ML potential (Fig. 1c) and its preceding validation for the even more disordered liquid (Fig. 1d) allow us to have complete confidence in the prediction. We note that McMillan *et al.* explicitly mention the presence of both polyamorphs on decompression, inferred from Raman data at the time;[9] further experimental investigations would certainly be worthwhile.

Upon even further compression, beginning at around 12 GPa, much higher coordinated ($N \geq 7$) regions suddenly emerged in our simulation (orange regions in Fig. 3b), again exhibiting spatial heterogeneity on a scale of several nanometres. These highly-coordinated regions rapidly coalesced into a new, dense form that is distinct from both LDA and HDA (Fig. 3c). We refer to this phase as "very-high-density amorphous" (VHDA), in line with conventions in the field.[14] The rapid structural collapse during VHDA formation reduced the volume from around 18 to around 14 Å$^3$ per atom (Fig. 3f). Importantly, this VHDA phase is transient in nature in these pressurization simulations, and crystalline regions rapidly nucleated in our simulation (Fig. 3d), in agreement with experiments: DAC X-ray measurements showed sharp diffraction peaks, consistent with an *sh* phase, beginning to appear at around 14 GPa.[11] The remarkable finding of the present work is not just the existence of *sh* at high pressure (that, alone, has been deduced from free-energy estimations[36], and observed by X-ray diffraction[11,12]), but the observation of a multistep crystallisation process which proceeds through an entirely distinct VHDA precursor phase – at variance with the assumption in previous work of direct HDA → crystalline transitions.[11,12,36] Having reached a pressure of 20 GPa (at the compression rate used, a few tens of picoseconds after the crystallisation had first set in), our system had fully transformed into a polycrystalline cell exhibiting hexagonally packed layers, stacked to form an *sh* structure (Fig. 3e and S3). Disordered regions between the crystal grains remained, as expected for poly- and nano-crystalline materials. This crystallisation mechanism cannot be described in conventional, small-scale simulations: we compressed a system containing 1,000 atoms (cell length ≈ 2.4 nm at 13 GPa) under otherwise similar conditions, and it retained a metastable VHDA-like structure up to 50 GPa (Fig. S5). Indeed, the small number of crystallites observed in our simulation (Fig. 3e) suggests a nucleation-controlled mechanism with a critical nucleus size of at least several atoms. It is challenging to quantify the critical nucleus size in these simulations, due to the highly disordered nature of the VHDA phase, but we may refer to an earlier, DFT-based thermodynamic estimate of a critical-nucleus diameter of ≈ 0.7 nm at 14 GPa,[11] much smaller than our simulation system size of > 10 nm for the 100,000-atom model. We note that an early DFT simulation[14] on a 216-atom *a*-Si model predicted an abrupt collapse of the tetrahedral network near 16 GPa (which we may now interpret as VHDA formation), though the tiny cell and short simulations revealed nothing about the stability of the structure, and did not show crystallization.[14] The pressure-induced crystallisation of amorphous solids appears to be a rather infrequent occurrence: two such instances include $Ge_2Sb_2Te_5$ (Ref. 37) and $Ce_{75}Al_{25}$,[38] but neither seem to involve (transient) VHDA-like phases in the crystallisation process.

Among the experimental indicators for the amorphous–amorphous transition in silicon is a sudden increase in the electrical conductivity.[9] Having accurate structural models available, we may study the electronic properties through the transition more closely, using tight-binding computations which are able to treat 100,000-atom size systems (Fig. 3h). A linear-scaling, maximum-entropy method[39] was combined with the tight-binding Hamiltonian of Kwon *et al.*,[40] previously used in studies of Urbach tails in $a$-Si[41] (Methods section). The density of states at the Fermi level, DOS($E_F$), plotted on the right-hand side of Fig. 3h, is a primary signature of electrical conductivity,[42] and its dramatic increase during compression indicates metallization in the transient VHDA phase, qualitatively consistent with the rapid conductivity increase between 10–12 GPa observed in DAC experiments.[9] At 13 GPa, the pseudogap is entirely filled in (left-hand side of Fig. 3h) and therefore DOS($E_F$) increases to a relative value of ≈ 17 (right-hand side). Interestingly, the computed DOS of the VHDA phase is virtually indistinguishable from that of the 1,500 K liquid at similar pressure (Fig. S7b), notwithstanding the temperature of our compression run (500 K), which is far below the melting line, and the somewhat larger degree of structural ordering of VHDA at lower temperature compared with that of the pressurised liquid at a higher temperature (Fig. S4). The prediction of this distinct electronic feature might be tested by ultrafast spectroscopy techniques, which have been previously applied to liquid silicon.[43] With even further increasing pressure and the initiation of crystallisation (> 15 GPa), a pseudogap appears again, reducing the normalised value of DOS($E_F$) to around 10. The pseudogap at $E_F$, seen at 20 GPa in Fig. 3h, may also be taken to indicate the stability of the polycrystalline $sh$-rich phase as compared to the transient VHDA structure.

The driving force for crystallisation can further be demonstrated by using, once more, the ML atomic energies (Fig. 3i). In the transient VHDA phase, the atomic-scale structural disorder is reflected in a seemingly random distribution of more stable (*blue*) and less stable (*yellow*) atomic environments. In contrast, the emerging $sh$ crystallites at 16 GPa provide distinct spatial regions of stability, around 0.3–0.5 eV at.$^{-1}$ above that of the fully relaxed diamond-type structure, in line with the computed equations of state for the $sh$ and diamond-type crystalline phases.[25] Our simulations have, therefore, described and explained the full range of phase transitions, reaching up to the experimentally established limit of existence (namely, $sh$ formation) of dense amorphous silicon.

Beyond this one specific material, our findings demonstrate the general ability of quantum-accurate, linear-scaling, ML-driven simulations to enable scientific discovery in a wide range of material systems. These methods can reveal yet unknown phases and phenomena, including those that emerge under extreme conditions, from individual atomic environments to nanoscale heterogeneity. Simulations of disordered materials have thereby taken a *qualitative* step forward: from simple structural models to predictive, realistic, and fully atomistic descriptions of material systems under experimentally challenging conditions.

# Methods

**An ML potential for dense disordered silicon.** Building on an existing general-purpose GAP for silicon,[25] henceforth referred to as "GAP-18", we created an extended version of this potential ("GAP-18P") that includes highly disordered high-pressure liquid and amorphous phases in the reference database. We adopted an iterative melt–quench (MQ) approach, as previously used for amorphous GeTe[44] and amorphous carbon.[31] In this, MQ simulations were run by a preliminary version of the potential and the resulting structures were evaluated by (single-point) DFT and fed back into the database. Here, to explore a wide range of pressures, we chose the unit-cell volume as a simple parameter, which we varied from 20 Å$^3$ at.$^{-1}$ (almost corresponding to ambient-pressure $a$-Si) down to 11 Å$^3$ at.$^{-1}$ (extreme compression). We performed GAP-driven constant-volume MQ simulations using a Langevin thermostat, as implemented in quippy (https://github.com/libAtoms/QUIP); the protocol is similar to that in Ref. 31. The reference DFT data were obtained using CASTEP 8.0 (Ref. 45) and the PW91 functional,[46] with convergence parameters as in the previous potential fit.[25]

The final potential uses the same fitting architecture as before, namely a baseline for exchange repulsion at short distances and a Smooth Overlap of Atomic Positions (SOAP)[33] descriptor and kernel for the GP regression. The SOAP parameters are a cut-off radius of 5.0 Å, a smoothness parameter of $\sigma_{at}$ = 0.5 Å, and a fit using 9,000 reference points. The unique identifier of the potential parameter files is `GAP_2018_5_14_60_21_25_5_692`. The potential can be run with LAMMPS[47] through the quippy interface (see below) and is freely available as Supplementary Information.

Test results for the new GAP-18P potential are given in Table S1, demonstrating that it does equally well for properties for which GAP-18 was initially certified. We computed the most central properties of crystalline (diamond-type) Si, including the optimised lattice parameter, $a$, the bulk modulus, $B$, and the formation energy of a single vacancy, $\Delta E_{vac}$. All these are within a few percent of the DFT reference and also deviate by no more than 1% from what is obtained with the general-purpose GAP-18 model.[25] We hence conclude that the application range of the new potential does not appear to be restricted compared to the previous formulation.

**Uncertainty quantification.** We evaluated the predicted error of the GAP models, which provides a measure for how far the potential is from configurations to which it has been fitted.[25] The GAP-18 was not fitted to high-pressure amorphous structures and therefore the prediction error for the VHDA structure at 13 GPa is correspondingly large (Fig. 1c). Our extended potential, GAP-18P, has encountered such configurations during fitting, and therefore its predicted error is low throughout (solid lines in Fig. 1c). It is notable that the 500 K snapshot at 20 GPa exhibits a bimodal distribution of errors when evaluated with GAP-18: most atoms in this structure are in crystalline-like (*sh*-like) environments, and such environments *are* included in the fitting database, leading to a low predicted error; a small number of atoms remains in disordered, interstitial regions (Fig. 3e) for which the predicted error is notably larger, centred at around 8 meV per atom. In contrast, GAP-18P shows a low predicted error throughout for this configuration, as well as for all others considered.

**Molecular-dynamics simulations.** MD simulations were carried out using LAMMPS,[47] with a Nosé–Hoover thermostat controlling temperature and a barostat controlling hydrostatic pressure.[48–50] The ambient-pressure quench follows the protocol established in our preceding pilot studies, and similarly uses the GAP-18 model: liquid Si at ambient pressure was quenched from 1,500 to 1,250 K at a rate of $10^{13}$ K s$^{-1}$, then to 1,050 K at $10^{11}$ K s$^{-1}$, and finally to 500 K at $10^{13}$ K s$^{-1}$. Pressurisation runs, using the GAP-18P model described above, were performed independently for the equilibrated 1,500 K liquid structure (pathway 1 in Fig. 1b; compressing to 14 GPa over 140 ps, and then to 23 GPa over another 40 ps) and for the quenched *a*-Si structure at 500 K (pathway 3; compressing to 20 GPa over 200 ps). The time step in all simulations was 1 fs.

**Electronic-structure computations.** The densities of electronic states (Fig. 3h) were obtained using the methods of Ref. 39. A relatively realistic tight-binding scheme using four orbitals (one s and three p) per site (Ref. 40) was employed to compute the Hamiltonian matrices for snapshots from 0 to 20 GPa, and also for large supercells of the diamond-type and simple hexagonal crystal phases of silicon. The electronic densities of states were computed with 70 Tchebychev polynomial moments extracted from sparse Hamiltonian matrices of dimension 400,000. For each snapshot, the 400,000 × 400,000 matrix was converted into a sparse format. A conservative initial guess, somewhat broader than the exact support of the spectrum, was made; the sparse Hamiltonian was then scaled and shifted onto the range (–1,1). An approximate "impartial vector" reproducing the first three exact moments was obtained,[39] and Tchebychev polynomial moments were extracted from the matrix [which, are in turn, Tchebychev moments of the density of states (DOS) function of the matrix]. The preceding matrix operations were order-$N$ ($N$ being the dimension of the matrix), since they required only matrix-on-vector operations[51] (no matrix multiplications). To obtain an approximate DOS, we solved the resulting Hausdorff moment problem. The Principle of Maximum Entropy[52] was used to solve the moment problem, both because of its underlying fundamental rationale, and its rapid pointwise convergence[53,54] compared to methods such as the Kernel Polynomial Method.[55] For large numbers of moments, numerical convergence is sensitive to the guessed spectral support, and this is iteratively tuned to the exact support as the number of moments increases. The convergence of the DOS was examined and 70 moments was found to be more than sufficient to obtain accurate pointwise estimates for the DOS across the full spectral range for all of our snapshots. In Fig. S7a, we illustrate the DOS computed with 70 and 120 moments for a 12 GPa snapshot: the functions are essentially identical. For reference, and to showcase the system sizes accessible to our method, we also include the DOS of the diamond-type structure (computed for a cubic 2,097,172-atom cell, 34.7584 nm on a side), using 170 moments, in Fig. S7c. This result may be compared to analogous computations in large fullerenes and graphene.[56]

**Acknowledgements**

V.L.D. acknowledges a Leverhulme Early Career Fellowship and support from the Isaac Newton Trust. Parts of the simulations reported here were carried out during his previous affiliation with the University of Cambridge (until August 2019). N.B. acknowledges support from the Office of Naval Research through the U.S. Naval Research Laboratory's core basic research program, and computer time through the U.S. DOD HPCMPO at the AFRL DSRC. D.A.D. acknowledges support from the U.S. NSF under award DMR 1506836. This work used the ARCHER UK National Supercomputing Service *via* a Resource Allocation Panel award (project e599) and the UKCP consortium (EPSRC grant EP/P022596/1). All structural drawings were created using OVITO.[57]

**Author contributions**

V.L.D., G.C., and S.R.E. initiated the project. V.L.D. and N.B. performed the ambient-pressure simulations; V.L.D. performed the high-pressure simulations; D.A.D. performed the electronic-structure computations. V.L.D., M.W., D.A.D., and S.R.E. analysed the data and developed the main conclusions regarding high-pressure phases. All authors contributed to discussions. V.L.D. created the figures and drafted the paper, and all authors contributed to its final version.

**Competing interests**

G.C. is listed as an inventor on a patent filed by Cambridge Enterprise Ltd. related to SOAP and GAP (PCT/GB2009/001414, filed on 5 June 2009 and published on 23 September 2014). The other authors declare no competing interests.

**Data availability**

Original data supporting this work, including coordinates for all reported structural models, will be made openly available through a suitable repository upon publication.

# References

1. Elliott, S. R. Medium-range structural order in covalent amorphous solids. *Nature* **354**, 445–452 (1991).

2. Sheng, H. W., Luo, W. K., Alamgir, F. M., Bai, J. M. & Ma, E. Atomic packing and short-to-medium-range order in metallic glasses. *Nature* **439**, 419–425 (2006).

3. Xie, R. *et al.* Hyperuniformity in amorphous silicon based on the measurement of the infinite-wavelength limit of the structure factor. *PNAS* **110**, 13250–13254 (2013).

4. Keen, D. A. & Goodwin, A. L. The crystallography of correlated disorder. *Nature* **521**, 303–309 (2015).

5. Hedler, A., Klaumünzer, S. L. & Wesch, W. Amorphous silicon exhibits a glass transition. *Nat. Mater.* **3**, 804–809 (2004).

6. Wilding, M. C., Wilson, M. & McMillan, P. F. Structural studies and polymorphism in amorphous solids and liquids at high pressure. *Chem. Soc. Rev.* **35**, 964 (2006).

7. Sheng, H. W. *et al.* Polyamorphism in a metallic glass. *Nat. Mater.* **6**, 192–197 (2007).

8. Deb, S. K., Wilding, M., Somayazulu, M. & McMillan, P. F. Pressure-induced amorphization and an amorphous–amorphous transition in densified porous silicon. *Nature* **414**, 528–530 (2001).

9. McMillan, P. F., Wilson, M., Daisenberger, D. & Machon, D. A density-driven phase transition between semiconducting and metallic polyamorphs of silicon. *Nat. Mater.* **4**, 680–684 (2005).

10. Daisenberger, D. *et al.* Polyamorphic amorphous silicon at high pressure: Raman and spatially resolved X-ray scattering and molecular dynamics studies. *J. Phys. Chem. B* **115**, 14246–14255 (2011).

11. Pandey, K. K., Garg, N., Shanavas, K. V., Sharma, S. M. & Sikka, S. K. Pressure induced crystallization in amorphous silicon. *J. Appl. Phys.* **109**, 113511 (2011).

12. Garg, N., Pandey, K. K., Shanavas, K. V., Betty, C. A. & Sharma, S. M. Memory effect in low-density amorphous silicon under pressure. *Phys. Rev. B* **83**, 115202 (2011).

13. Haberl, B., Guthrie, M., Sprouster, D. J., Williams, J. S. & Bradby, J. E. New insight into pressure-induced phase transitions of amorphous silicon: the role of impurities. *J. Appl. Crystallogr.* **46**, 758–768 (2013).

14. Durandurdu, M. & Drabold, D. A. *Ab initio* simulation of first-order amorphous-to-amorphous phase transition of silicon. *Phys. Rev. B* **64**, 014101 (2001).

15. Morishita, T. High density amorphous form and polyamorphic transformations of silicon. *Phys. Rev. Lett.* **93**, 055503 (2004).

16. Daisenberger, D. *et al.* High-pressure x-ray scattering and computer simulation studies of density-induced polyamorphism in silicon. *Phys. Rev. B* **75**, 224118 (2007).

17. Mishima, O., Calvert, L. D. & Whalley, E. An apparently first-order transition between two amorphous phases of ice induced by pressure. *Nature* **314**, 76–78 (1985).

18. Mishima, O., Takemura, K. & Aoki, K. Visual observations of the amorphous-amorphous transition in $H_2O$ under pressure. *Science* **254**, 406–408 (1991).

19. Behler, J. First principles neural network potentials for reactive simulations of large molecular and condensed systems. *Angew. Chem. Int. Ed.* **56**, 12828–12840 (2017).

20. Butler, K. T., Davies, D. W., Cartwright, H., Isayev, O. & Walsh, A. Machine learning for molecular and materials science. *Nature* **559**, 547–555 (2018).

21. Deringer, V. L., Caro, M. A. & Csányi, G. Machine learning interatomic potentials as emerging tools for materials science. *Adv. Mater.* **31**, 1902765 (2019).

22. Behler, J., Martoňák, R., Donadio, D. & Parrinello, M. Metadynamics simulations of the high-pressure phases of silicon employing a high-dimensional neural network potential. *Phys. Rev. Lett.* **100**, 185501 (2008).

23. Bonati, L. & Parrinello, M. Silicon liquid structure and crystal nucleation from *ab initio* deep metadynamics. *Phys. Rev. Lett.* **121**, 265701 (2018).

24. Bartók, A. P., Payne, M. C., Kondor, R. & Csányi, G. Gaussian approximation potentials: The accuracy of quantum mechanics, without the electrons. *Phys. Rev. Lett.* **104**, 136403 (2010).

25. Bartók, A. P., Kermode, J., Bernstein, N. & Csányi, G. Machine learning a general-purpose interatomic potential for silicon. *Phys. Rev. X* **8**, 041048 (2018).

26. Deringer, V. L. *et al.* Realistic atomistic structure of amorphous silicon from machine-learning-driven molecular dynamics. *J. Phys. Chem. Lett.* **9**, 2879–2885 (2018).

27. Bernstein, N. *et al.* Quantifying chemical structure and machine-learned atomic energies in amorphous and liquid silicon. *Angew. Chem. Int. Ed.* **58**, 7057–7061 (2019).

28. Hejna, M., Steinhardt, P. J. & Torquato, S. Nearly hyperuniform network models of amorphous silicon. *Phys. Rev. B* **87**, 245204 (2013).

29. Dahal, D., Atta-Fynn, R., Elliott, S. R. & Biswas, P. Hyperuniformity and static structure factor of amorphous silicon in the infinite-wavelength limit. *J. Phys.: Conf. Ser.* **1252**, 012003 (2019).

30. Oganov, A. R., Pickard, C. J., Zhu, Q. & Needs, R. J. Structure prediction drives materials discovery. *Nat. Rev. Mater.* **4**, 331–348 (2019).

31. Deringer, V. L. & Csányi, G. Machine learning based interatomic potential for amorphous carbon. *Phys. Rev. B* **95**, 094203 (2017).

32. Funamori, N. & Tsuji, K. Pressure-induced structural change of liquid silicon. *Phys. Rev. Lett.* **88**, 255508 (2002).

33. Bartók, A. P., Kondor, R. & Csányi, G. On representing chemical environments. *Phys. Rev. B* **87**, 184115 (2013).

34. De, S., Bartók, A. P., Csányi, G. & Ceriotti, M. Comparing molecules and solids across structural and alchemical space. *Phys. Chem. Chem. Phys.* **18**, 13754–13769 (2016).

35. Imai, M., Mitamura, T., Yaoita, K. & Tsuji, K. Pressure-induced phase transition of crystalline and amorphous silicon and germanium at low temperatures. *High Pressure Res.* **15**, 167–189 (1996).

36. Shanavas, K. V., Pandey, K. K., Garg, N. & Sharma, S. M. Computer simulations of crystallization kinetics in amorphous silicon under pressure. *J. Appl. Phys.* **111**, 063509 (2012).

37. Xu, M. *et al.* Pressure-induced crystallization of amorphous $Ge_2Sb_2Te_5$. *J. Appl. Phys.* **108**, 083519 (2010).

38. Wu, M., Tse, J. S., Wang, S. Y., Wang, C. Z. & Jiang, J. Z. Origin of pressure-induced crystallization of $Ce_{75}Al_{25}$ metallic glass. *Nat. Commun.* **6**, 6493 (2015).

39. Drabold, D. A. & Sankey, O. F. Maximum entropy approach for linear scaling in the electronic structure problem. *Phys. Rev. Lett.* **70**, 3631–3634 (1993).

40. Kwon, I., Biswas, R., Wang, C. Z., Ho, K. M. & Soukoulis, C. M. Transferable tight-binding models for silicon. *Phys. Rev. B* **49**, 7242–7250 (1994).

41. Drabold, D. A., Li, Y., Cai, B. & Zhang, M. Urbach tails of amorphous silicon. *Phys. Rev. B* **83**, 045201 (2011).

42. Mott, N. F. & Davis, E. A. *Electronic Processes in Non-Crystalline Materials*. (Oxford University Press, 2012).

43. Beye, M., Sorgenfrei, F., Schlotter, W. F., Wurth, W. & Fohlisch, A. The liquid-liquid phase transition in silicon revealed by snapshots of valence electrons. *Proc. Natl. Acad. Sci., U. S. A.* **107**, 16772–16776 (2010).

44. Sosso, G. C., Miceli, G., Caravati, S., Behler, J. & Bernasconi, M. Neural network interatomic potential for the phase change material GeTe. *Phys. Rev. B* **85**, 174103 (2012).

45. Clark, S. J. *et al.* First principles methods using CASTEP. *Z. Krist.* **220**, 567–570 (2005).

46. Perdew, J. P. & Wang, Y. Accurate and simple analytic representation of the electron-gas correlation energy. *Phys. Rev. B* **45**, 13244–13249 (1992).

47. Plimpton, S. Fast parallel algorithms for short-range molecular dynamics. *J. Comput. Phys.* **117**, 1–19 (1995).

48. Parrinello, M. & Rahman, A. Polymorphic transitions in single crystals: A new molecular dynamics method. *J. Appl. Phys.* **52**, 7182–7190 (1981).

49. Martyna, G. J., Tobias, D. J. & Klein, M. L. Constant pressure molecular dynamics algorithms. *J. Chem. Phys.* **101**, 4177–4189 (1994).

50. Shinoda, W., Shiga, M. & Mikami, M. Rapid estimation of elastic constants by molecular dynamics simulation under constant stress. *Phys. Rev. B* **69**, 134103 (2004).

51. Skilling, J. in *Maximum Entropy and Bayesian Methods* (ed. Skilling, J.) (Kluwer, 1989).

52. Jaynes, E. T. *Probability Theory: the Logic of Science*. (Cambridge University Press, 2003).

53. Mead, L. R. & Papanicolaou, N. Maximum entropy in the problem of moments. *J. Math. Phys.* **25**, 2404–2417 (1984).

54. Bandyopadhyay, K., Bhattacharya, A. K., Biswas, P. & Drabold, D. A. Maximum entropy and the problem of moments: A stable algorithm. *Phys. Rev. E* **71**, 057701 (2005).

55. Weiße, A., Wellein, G., Alvermann, A. & Fehske, H. The kernel polynomial method. *Rev. Mod. Phys.* **78**, 275–306 (2006).

56. Drabold, D. A., Ordejón, P., Dong, J. & Martin, R. M. Spectral properties of large fullerenes: From cluster to crystal. *Sol. State Commun.* **96**, 833–838 (1995).

57. Stukowski, A. Visualization and analysis of atomistic simulation data with OVITO–the Open Visualization Tool. *Model. Simul. Mater. Sci. Eng.* **18**, 015012 (2010).

58. Laaziri, K. *et al.* High-energy x-ray diffraction study of pure amorphous silicon. *Phys. Rev. B* **60**, 13520–13533 (1999).

# Supplementary Material

**Table S1:** Fundamental properties of diamond-type crystalline silicon, obtained with the previous general-purpose potential ("GAP-18", Ref. 25) and the extended potential ("GAP-18P") developed in the present study. Specifically, we evaluate the lattice parameter, $a$, the bulk modulus, $B$, the elastic constants, $c_{11}$, $c_{12}$, and $c_{44}$, and the formation energy of a vacancy, $\Delta E_{\mathrm{vac}}$. We also compute the energies of various relaxed ($hkl$) surfaces, including important ($n\times m$) reconstructions, and finally the energy required to form the (112) $\Sigma 3$ grain boundary (GB), relative to the ideal diamond structure. Sets of results obtained with both GAPs are compared to DFT results at the same level (relative errors are given for the GAPs, rounded to full integer numbers).

|  | DFT | GAP-18 |  | GAP-18P |  |
|---|---|---|---|---|---|
| *Bulk properties* | | | | | |
| $a$ (Å) | 5.461 | 5.461 | (±0%) | 5.467 | (< 1%) |
| $B$ (GPa) | 88.6 | 88.6 | (±0%) | 88.6 | (±0%) |
| $c_{11}$ (GPa) | 153.3 | 149.2 | (–3%) | 149.5 | (–2%) |
| $c_{12}$ (GPa) | 56.3 | 58.2 | (+4%) | 58.1 | (+3%) |
| $c_{44}$ (GPa) | 72.2 | 66.7 | (–8%) | 66.8 | (–8%) |
| $\Delta E_{\mathrm{vac}}$ (eV) | 3.67 | 3.60 | (–2%) | 3.61 | (–2%) |
| *Surface and grain boundary energies (all in meV Å$^{-2}$)* | | | | | |
| (001), (1×1) | 136 | 133 | (–2%) | 133 | (–2%) |
| (001), (2×1) | 82 | 86 | (+5%) | 85 | (+5%) |
| (111), (1×1) | 98 | 96 | (–2%) | 96 | (–2%) |
| (111), (7×7) | 86 | 84 | (–2%) | 84 | (–2%) |
| GB (112) $\Sigma 3$ | 58 | 60 | (+3%) | 59 | (+1%) |

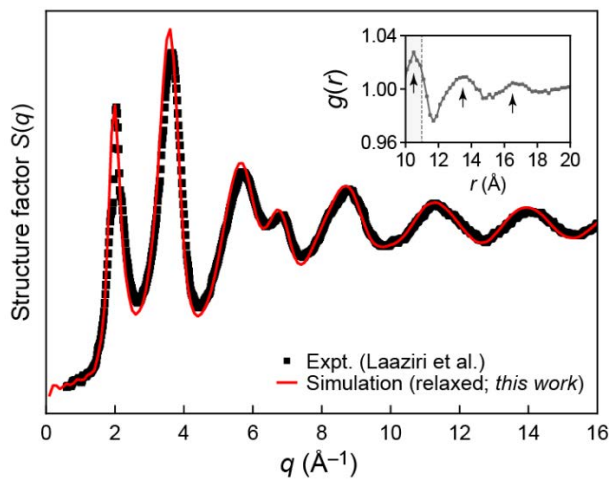

**Figure S1: Computed structure factor for the fully optimised *a*-Si structure.** The static structure factor, $S(q)$, as a probe for medium-range structural order, has been evaluated for the relaxed amorphous system in Fig. 1a and compared to experimental data from Ref. 58. The inset shows a radial distribution function, $g(r)$, for the same structure, indicating long-range correlations beyond the first nanometre, which our ML-driven simulations can access. A dashed line at $\approx 11$ Å indicates the limit of DFT modelling, corresponding to half the cell length of the smallest system sketched in Fig. 1a.

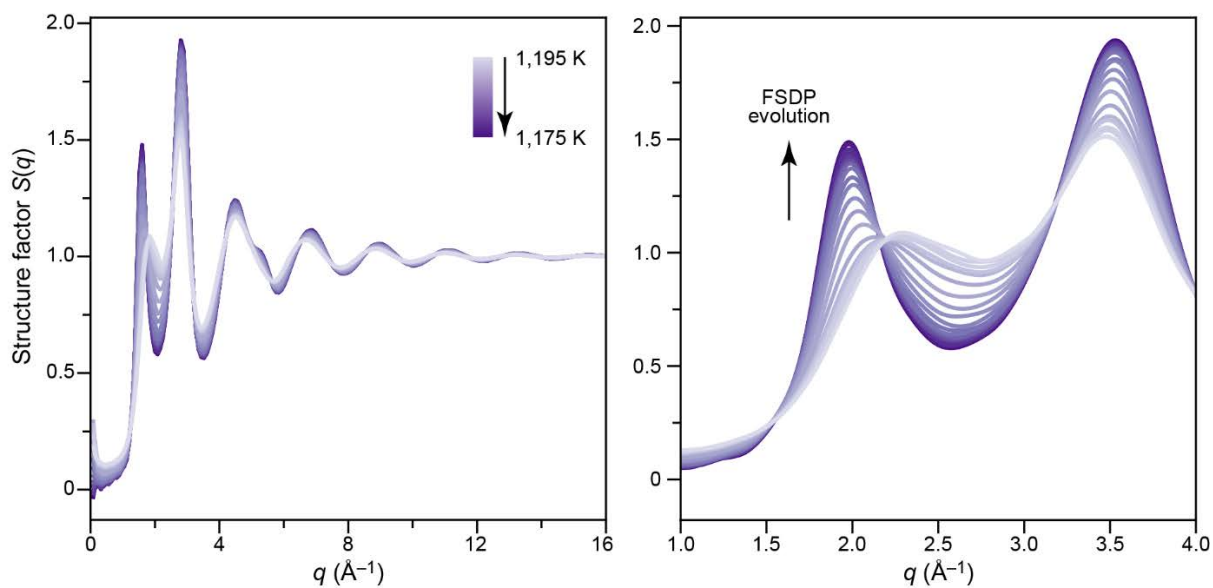

**Figure S2: Computed structure factors during quenching.** The plot on the left-hand side shows the evolution of simulated structure factors through the relevant part of the liquid-quenching trajectory (pathway 2) in the vicinity of the glass transition, plotted in 1 K temperature increments. The emergence of the first sharp diffraction peak, FSDP (between 1.5 and 2.0 Å$^{-1}$), as well as the structuring of the third peak (between 5 and 6 Å$^{-1}$), are clearly visible. On the right-hand side, a detailed view is provided of the evolution of the FSDP.

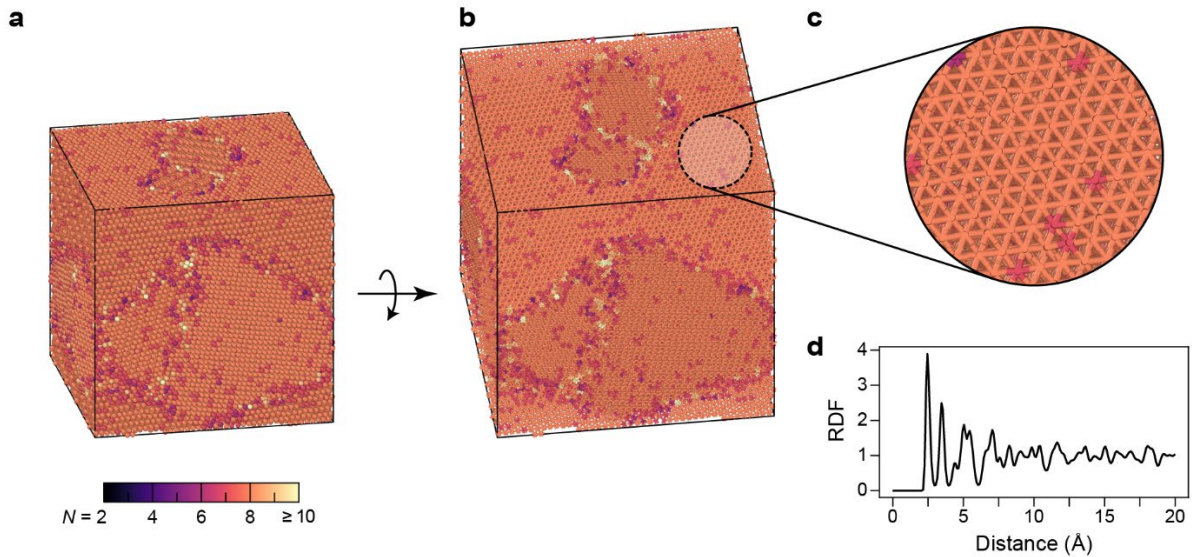

**Figure S3: Crystallisation of *a*-Si under high pressure.** We show the compressed 100,000-atom system at 20 GPa, emphasising its poly-crystalline nature. Two slightly different views of the structure are provided, representing atoms (spheres; panel **a**) or bonds up to a spatial cut-off of 2.85 Å (cylinders; panel **b**). The second structure has been slightly rotated to show the presence of 2D hexagonal packing, characteristic of the simple hexagonal structure, shown as a close-up in panel (**c**). The computed radial distribution function (panel **d**) furthermore emphasises the structural ordering in the poly-crystalline system.

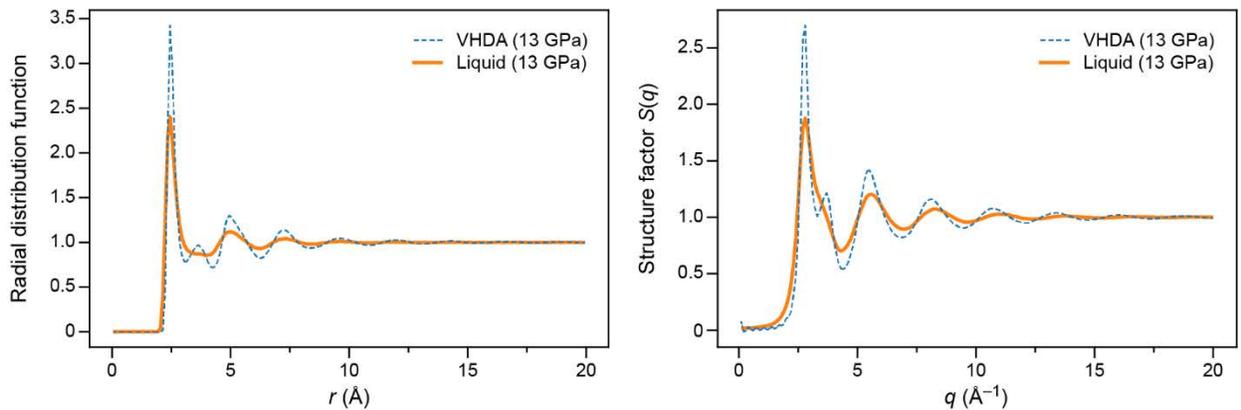

**Figure S4: Structural features of VHDA and the high-pressure liquid.** We analyse two snapshots at the same pressure (13 GPa, corresponding to Fig. 3c in the main text, where the VHDA phase is seen), but necessarily at different temperatures (500 K for VHDA, 1,500 K for the liquid), showing computed radial distribution functions (*left-hand side*) and structure factors (*right-hand side*) to emphasise structural similarities and differences between the two phases.

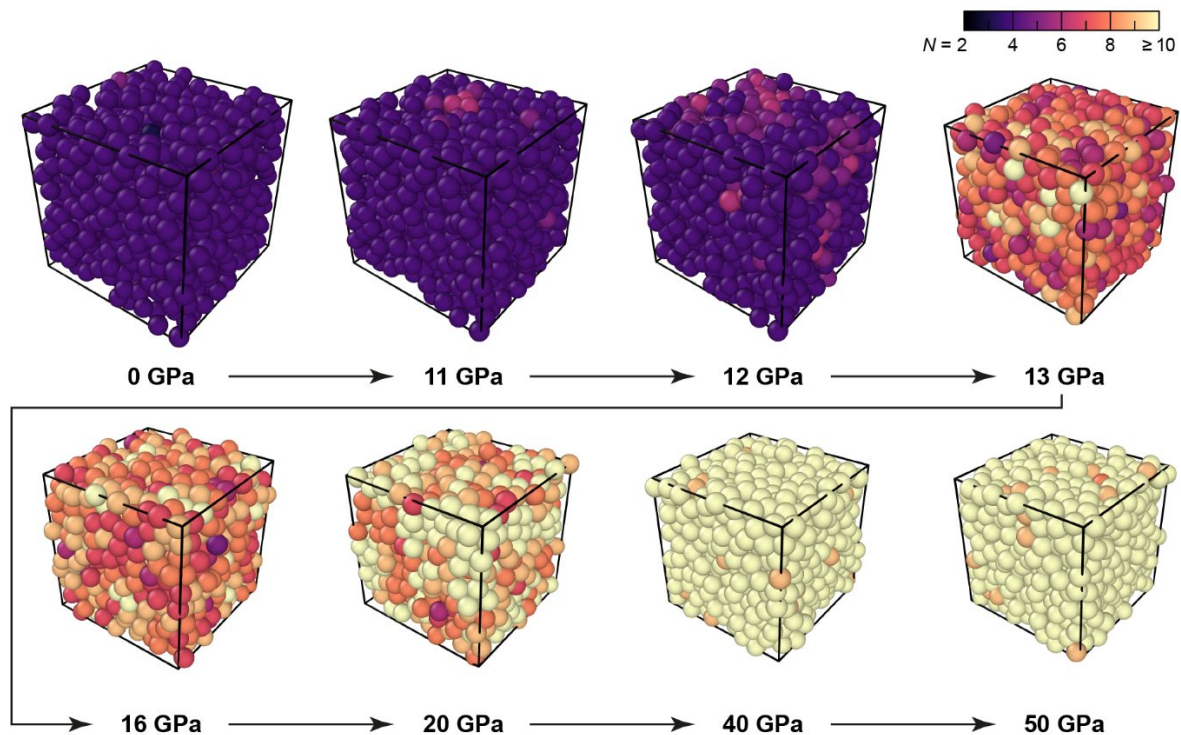

**Figure S5: Compressing a 1,000-atom *a*-Si structure.** In this case, despite an otherwise similar simulation protocol (and using the GAP-18P model throughout), no crystallisation was observed; instead, the VHDA phase densified further upon compression up to 50 GPa. Further compression runs were performed with 1,000-atom *a*-Si structures that had been quenched at constant external pressures of 1, 2, and 5 GPa, respectively; only the last of them led to crystallisation, whereas the others did not, and we therefore conclude that a system size of 1,000 atoms (cell length ~2.4 nm at 13 GPa) will be too reliant on stochastic effects (*e.g.*, the existence of a crystallisation nucleus) to describe the multiscale physical nature of the system correctly.

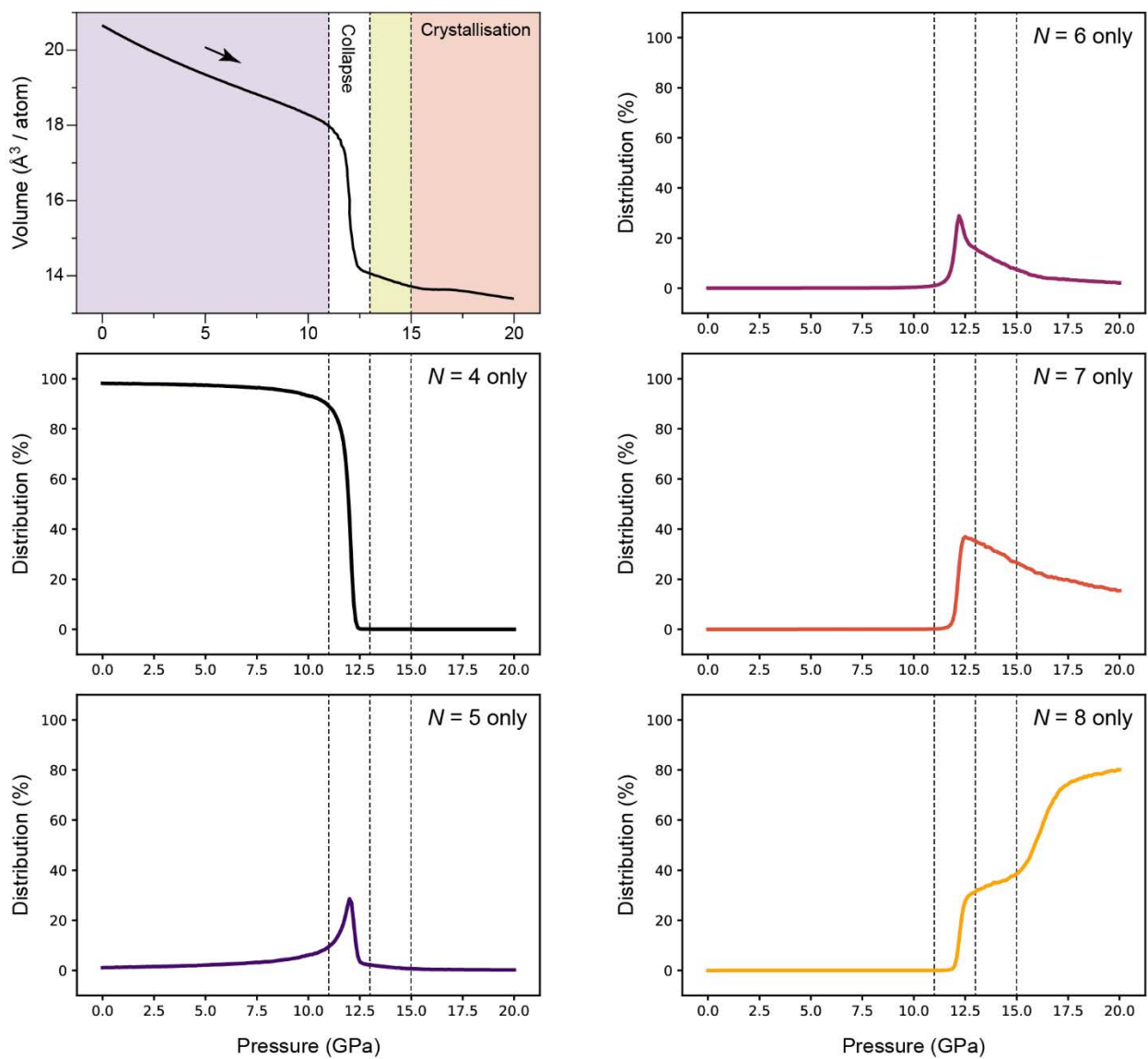

**Figure S6** *(as a supplement to Fig. 3f in the main text)*: **Evolution of coordination numbers for the 100,000-atom model during compression.** This figure provides a more detailed representation of the data shown in the inset of Fig. 3f, *i.e.*, the abundance of given coordination numbers, $N$, during the full compression run (coordination numbers determined using a 2.85 Å spatial cut-off). The volume-versus-pressure plot is reproduced from the main text for reference.

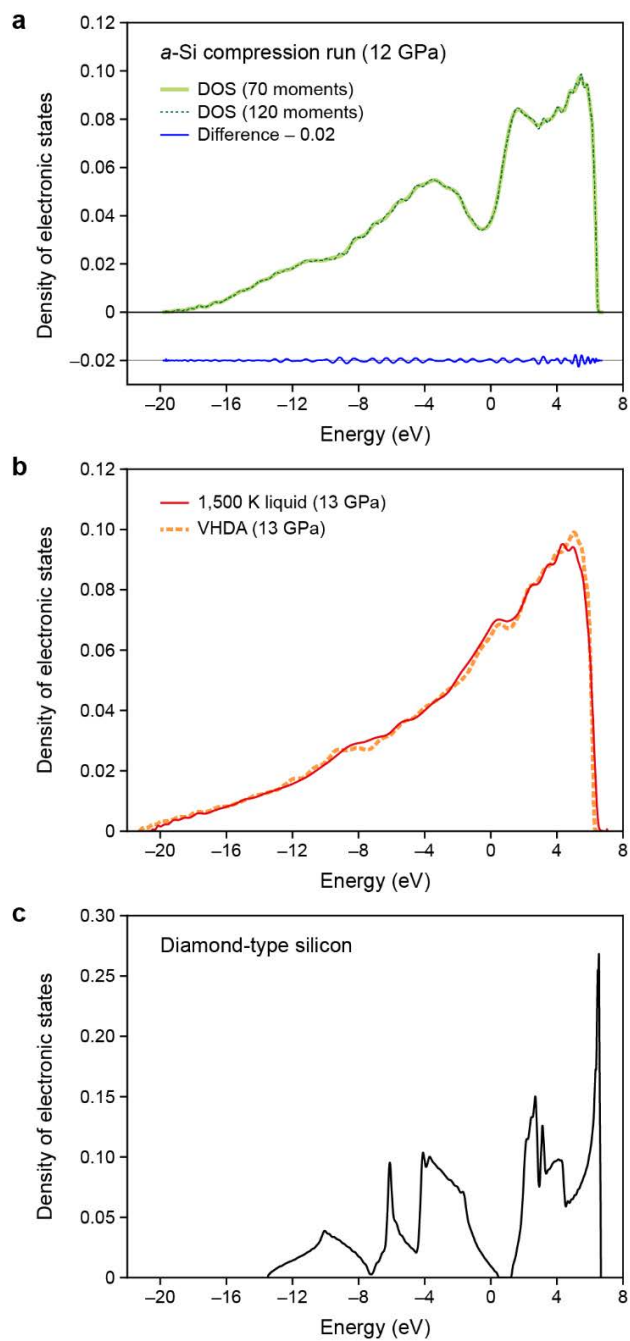

**Figure S7** *(as a supplement to Fig. 3h in the main text)*: **Additional results for electronic-structure computations on ultra-large systems.** Panel (**a**) compares the electronic DOS computed using 70 moments (*solid orange line*) and 120 moments (*dashed red line*), which are practically indistinguishable; a difference plot (*blue*) is given below (shifted vertically by –0.02). Panel (**b**) compares the electronic structure of the transient VHDA phase to that of the 1,500 K liquid at the same pressure. The similarity in the DOS curves reflects the structural similarity of the two states (Fig. S4). Panel (**c**) shows the electronic DOS computed with the same approach but for a diamond-type crystalline silicon supercell at atmospheric pressure, and containing > 2 million atoms (see details in the Methods section). The energy scale is set by Ref. 40.